\documentclass[aps,prb,nofootinbib,twocolumn,floatfix,groupedaddress,superscriptaddress]{revtex4}

\usepackage{latexsym} 
\usepackage{amsmath} 
\usepackage{dcolumn}
\usepackage{bm} 
\usepackage{graphicx} 
\usepackage{epsfig}
\usepackage{float}
\usepackage[]{graphicx}
\usepackage[]{wrapfig}
\usepackage[]{color}

\begin{document}

\preprint{APS/123-QED}
\title{A real-time TDDFT study of femtosecond laser-driven monolayer NbSe$_2$}

\author{Towfiq Ahmed}
\email{atowfiq@lanl.gov}
\affiliation{Theoretical Division, Los Alamos National Laboratory, Los Alamos, New Mexico 87545, USA}

\author{Jinkyoung Yoo}
\affiliation{Center for Integrated Nanotechnologies, Los Alamos National Laboratory, Los Alamos, New Mexico 87545, USA}
    
\author{Rohit Prasankumar}
\affiliation{Center for Integrated Nanotechnologies, Los Alamos National Laboratory, Los Alamos, New Mexico 87545, USA}

\author{Jian-Xin Zhu}
\affiliation{Theoretical Division, Los Alamos National Laboratory, Los Alamos, New Mexico 87545, USA}
\affiliation{Center for Integrated Nanotechnologies, Los Alamos National Laboratory, Los Alamos, New Mexico 87545, USA}

\date{\today}

\begin{abstract}
High harmonic generation (HHG) spectra have the potential to show novel signatures of ordered phases in condensed matter. We studied the femtosecond laser-driven electronic response of monolayer NbSe$_2$ using state-of-the-art computational methods, which can guide future synthesis and optical characterization. Earlier studies found distinct signatures of charge density wave (CDW) ordered phases in the ground state of NbSe$_2$ monolayers, in co-existence with superconductivity. Driving such systems with ultrashort laser pulses can shed new light on optically controlling various exotic phases (e.g. CDW) in monolayer NbSe$_2$. This will not only provide a fundamental understanding of non-equilibrium phase-transitions in NbSe$_2$, but also will open a path forward for revolutionizing quantum information technologies, such as valleytronics. To this end, we have studied high harmonic generation (HHG) in monolayer NbSe$_2$ under various optical pump intensities using real-time time-dependent density functional theory (RT-TDDFT). Our calculations predict distinct signatures in HHG spectra for certain harmonics in the presence of CDW order in monolayer NbSe$_2$. Finally, we also examined the dependence of HHG spectra on excitation intensity and qualitatively revealed its power-law behavior.
\end{abstract}


\maketitle

\section{\label{sec:level1}Introduction}
Electronic instabilities have led to novel phases such as superconductivity (SC) and charge density waves (CDW) in several low dimensional transition-metal dichalcogenide (TMDC) systems. Extensive experimental and theoretical studies have identified SC phases in layered PdTe$_2$~\cite{2D_PdTe2}, 2H-TaS$_2$~\cite{tmdc_2b}, 
2H-TaSe$_2$~\cite{tmdc_2}, and 2H-NbSe$_2$~\cite{tmdc_1,tmdc_3a}. Interestingly, a coexisting CDW phase has also been identified in 
2H-NbSe$_2$~\cite{cdw_4,cdw_3} below $T_c^{CDW}$ = 33 K ( 145 K in monolayer~\cite{synthesis_1}), and in 2H-TaSe$_2$\cite{cdw_1a} below $T_c^{CDW}$ = 120 K. Earlier neutron scattering~\cite{NbSe2_1} and X-ray diffraction~\cite{NbSe2_3} data have revealed the existence of a periodic lattice distortion in monolayer NbSe$_2$, which causes the CDW ordering. However, the fundamental mechanisms governing the nature of the CDW phase in monolayer NbSe$_2$ have not been fully understood and remain a matter of great interest~\cite{cdw_4}. Although a two-dimensional (2D) Peierls-type instability due to Fermi surface nesting was proposed earlier~\cite{FSnesting}, more compelling experimental evidence and theoretical arguments are now in favor of electron-phonon-mediated CDW order formation in monolayer NbSe$_2$,~\cite{epc_1a,epc_2a,epc_1,epc_2} which remains metallic below 145 K and superconducting below 1 K.\cite{synthesis_1} Realization of such unique metallic CDW phase in NbSe$_2$ has been supported by structural characterization~\cite{NbSe2_1,NbSe2_3,NbSe2_4}, where slightly incommensurate CDW modulation in a $3\times3$ supercell of monolayer NbSe$_2$ was observed. 
\begin{figure}
 \includegraphics[scale=0.33,angle=0]{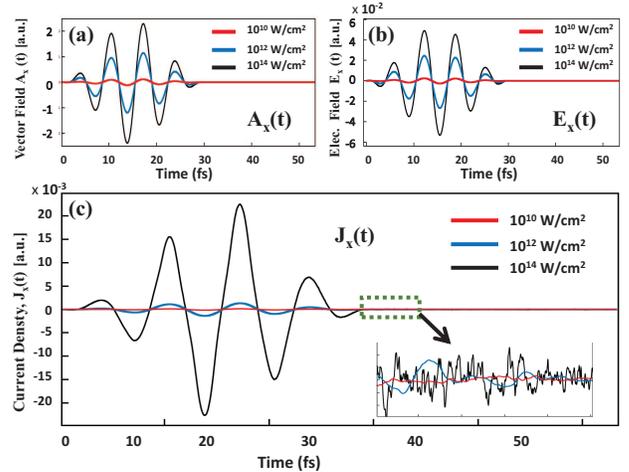}
  \caption
   {(Color online) Femtosecond laser pulse and current density in an undistorted pristine NbSe$_2$ monolayer. (a) Incident external pulse field in the time domain with different intensities: 10$^{10}$ (red), 10$^{12}$ (blue), and 10$^{14}$ (black) W/cm$^2$. 
   (b) Fourier transformed pulse field in the frequency domain for an intensity of 10$^{14}$ W/cm$^2$.
   The real and imaginary parts of the frequency-dependent field are shown with black and blue curves, respectively. (c) Current density temporal response of the system under different intensities. The inset shows finer spectral structure that was enhanced with higher intensity pulses.
}\label{fig1}
\end{figure}

In this Article, we study the electronic response of the CDW ordered phase of monolayer NbSe$_2$ under femtosecond optical excitation. Electronic excitation under a strong optical driving field creates rich phenomena, such as high harmonic generation (HHG)~\cite{Ghimire2019,HHG_Brabec}, that can change across a non-equilibrium phase transition. HHG spectra of solids often carry valuable spectral signatures, which can help identify dynamic structural symmetry-breaking (or ordering) ~\cite{boyd}. Controlling such non-equilibrium phases in monolayer NbSe$_2$ with femtosecond laser pulses could, in principle, provide a pathway for understanding the fundamental competition between order parameters (e.g. SC and CDW). This in turn will potentially lead to new insights for tuning material functionalities, such as conductance and valley degrees of freedom~\cite{valleyP}, that are highly relevant for quantum information technology.

Here, to theoretically examine the CDW ordered phase of monolayer NbSe$_2$ under an intense optical field, we employed real-time time-dependent density functional theory 
(RT-TDDFT)~\cite{gross,ullrich} as implemented in the SALMON code~\cite{salmon_1}. A state-of-the-art non-perturbative TDDFT formulation, also known as the {\it velocity gauge}~\cite{Yabana_1,velocity_gauge,HHG_Lu,ShengMeng2019} method, was used in our theoretical approach in order to preserve translation symmetry in periodic solids under a strong driving field. Starting with an undistorted (non-CDW) monolayer of 2H-NbSe$_2$, we calculated HHG spectra for varying intensities. Several peak features (odd and even higher harmonic modes) displayed the qualitative nature of symmetry in the system, while a nonlinear electronic response was identified with increasing field amplitude. In this nonlinear limit, we then studied the HHG spectra of a CDW ordered  NbSe$_2$ monolayer by considering a commensurate $3\times3$ lattice. Earlier theoretical work~\cite{NbSe2_2} found negligible difference in the electronic structure between commensurate and nearly incommensurate  superlattice structures in 2D NbSe$_2$. Here, our calculations show interesting differences in some particular HHG spectral modes between CDW and non-CDW-ordered phases in NbSe$_2$ monolayers, thus providing guidance for the experimental identification of structural distortions associated with CDW ordering under intense optical excitation.

\section{\label{sec:level2}Methods} 
First-principles RT-TDDFT calculations were performed to understand the nonlinear response of monolayer NbSe$_2$ under a strong 
optical field. 

\subsection{\label{sec:level2a}Theoretical Approach}
The workhorse of our calculations is RT-TDDFT, which is a non-perturbative method suitably designed for electronic excitations under a strong external field. This real-time approach is developed on the foundation of conventional density functional theory (DFT)~\cite{dft_1,dft_2}, but goes well beyond the perturbative or linear response range of the time-dependent DFT (TDDFT)~\cite{tddft_1} approach. The two key features on which the RT-TDDFT method for periodic systems relies are (i) stable and efficient real-time propagation of the time-dependent Kohn-Sham orbitals using a time-dependent Hamiltonian, and (ii) `velocity gauge' formulation of the Hamiltonian, where a time-dependent vector gauge field is used in the kinetic term instead of a time-dependent interaction term, which otherwise would break the translation symmetry of periodic systems.

Bertsch and Yabana~\cite{Yabana_1,velocity_gauge} originally proposed the velocity gauge formulation of the real-time time-dependent Kohn-Sham (TDKS) equation, which has recently been implemented in a few TDDFT codes. We briefly describe the governing RT-TDDFT equations here, starting with a general time-dependent Kohn-Sham Hamiltonian: 
\begin{eqnarray}
\label{eqn:eqn_01}
    \it{i}\hbar\frac{\partial}{\partial t}\psi_i(\textbf{r},t)=\hat{H}_{KS} \psi_i(\textbf{r},t),
\end{eqnarray}
where 
\begin{eqnarray}
\label{eqn:eqn_02}
 \hat{H}_{KS} = \frac{\textbf{p}^2}{2 m} &+& \hat{V}_{ion} + \int d\textbf{r}' \frac{e^2}{|\textbf{r} - \textbf{r}'|} n(\textbf{r},t) \nonumber \\
 &+& V_{xc}[n(\textbf{r},t)] + e \textbf{E}(t) \cdot \textbf{r}. 
 \end{eqnarray}
 The third term represents the Hartree Coulomb interaction, while the fourth term $V_{xc}$ is the exchange-correlation potential and 
 the electron density $n(\textbf{r},t) = \sum_i |\psi(\textbf{r},t)|^2$. 
 One instantly recognizes that the time-dependent external interaction potential term
 $e \textbf{E}(t) \cdot \bf {r}$ breaks the
 translational symmetry of periodic systems. Such concerns are addressed by adopting the `velocity gauge', where a vector field, defined as
\begin{eqnarray}
\label{eqn:eqn_03}
\textbf{A}(t) = -c\int^t \textbf{E}(t') dt',
\end{eqnarray} 
is used to gauge-transform the KS wave functions as
\begin{eqnarray}
\label{eqn:eqn_04}
\psi^\prime(\textbf{r},t) = \exp\left[\frac{ie}{\hbar c} \textbf{A}(t) \cdot \textbf{r}\right] \psi(\textbf{r},t).
\end{eqnarray} 
The velocity-gauge TDKS Hamiltonian now takes the form:
\begin{eqnarray}
\label{eqn:eqn_05}
 \hat{H}^{RT}_{KS} &=& \frac{1}{2m}\left[\textbf{p} + \frac{e}{c}\textbf{A}(t)\right]^2 + \hat{V}_{ion} \nonumber \\
 &+& \int d\textbf{r}' \frac{e^2}{|\textbf{r} - \textbf{r}'|} n(\textbf{r},t) + V_{xc}[n(\textbf{r},t)], \end{eqnarray} 
 The external field or interaction potential is now incorporated in the kinetic energy term, and consequently the translational symmetry of the Hamiltonian is restored. The time-dependent KS orbitals are now evolved using:
 \begin{eqnarray}
\label{eqn:eqn_06}
    \it{i}\hbar\frac{\partial}{\partial t}\psi_i(\textbf{r},t)=\hat{H}^{RT}_{KS} \psi_i(\textbf{r},t).
\end{eqnarray}
Different time evolution propagators are constructed from the above equation, and their forms depend on the choice of basis set (e.g. overlap and Hamiltonian matrix elements) of the original DFT formalism. Detailed discussion of the propagators can be found elsewhere~\cite{rttddft_hubbard,propagator}.

Finally, the time-dependent current density 
\begin{eqnarray}
\label{eqn:eqn_05a}
\textbf{J}(t) = -\frac{i}{2\Omega}\int_{\Omega} d\textbf{r} \sum_i \left[ \psi^*_i(\textbf{r},t)\nabla \psi_i(\textbf{r},t) - \text{c.c.}\right]
\end{eqnarray} 
is obtained, whose Fourier transform gives the HHG spectra:
\begin{eqnarray}
\label{eqn:eqn_06a}
\textbf{HHG}(\omega)=\omega^2\left|\int^T_0 \textbf{J}(t)\exp(-i\omega t)dt\right|^2.
\end{eqnarray} 

One of the central features of the real-time TDDFT method is its real-time propagators, which enables one to treat the external pulse field with high intensity in a non-perturbative manner, as opposed to frequency-dependent TDDFT with linear response implementation. However, one must propagate the system long enough to capture the system's response under the the external field. Therefore, such computations are mostly suitable for short pulse fields. On the other hand, for a time-periodic external field, one may consider the Floquet method, which is most efficiently solved in frequency space. The Floquet treatment in a real-time approach may not be feasible, given the computational cost. The system needs to be time-propagated for a very long time in order to capture the nonlinear response under a time-periodic external field. 

\begin{figure}
 \includegraphics[scale=0.33,angle=0]{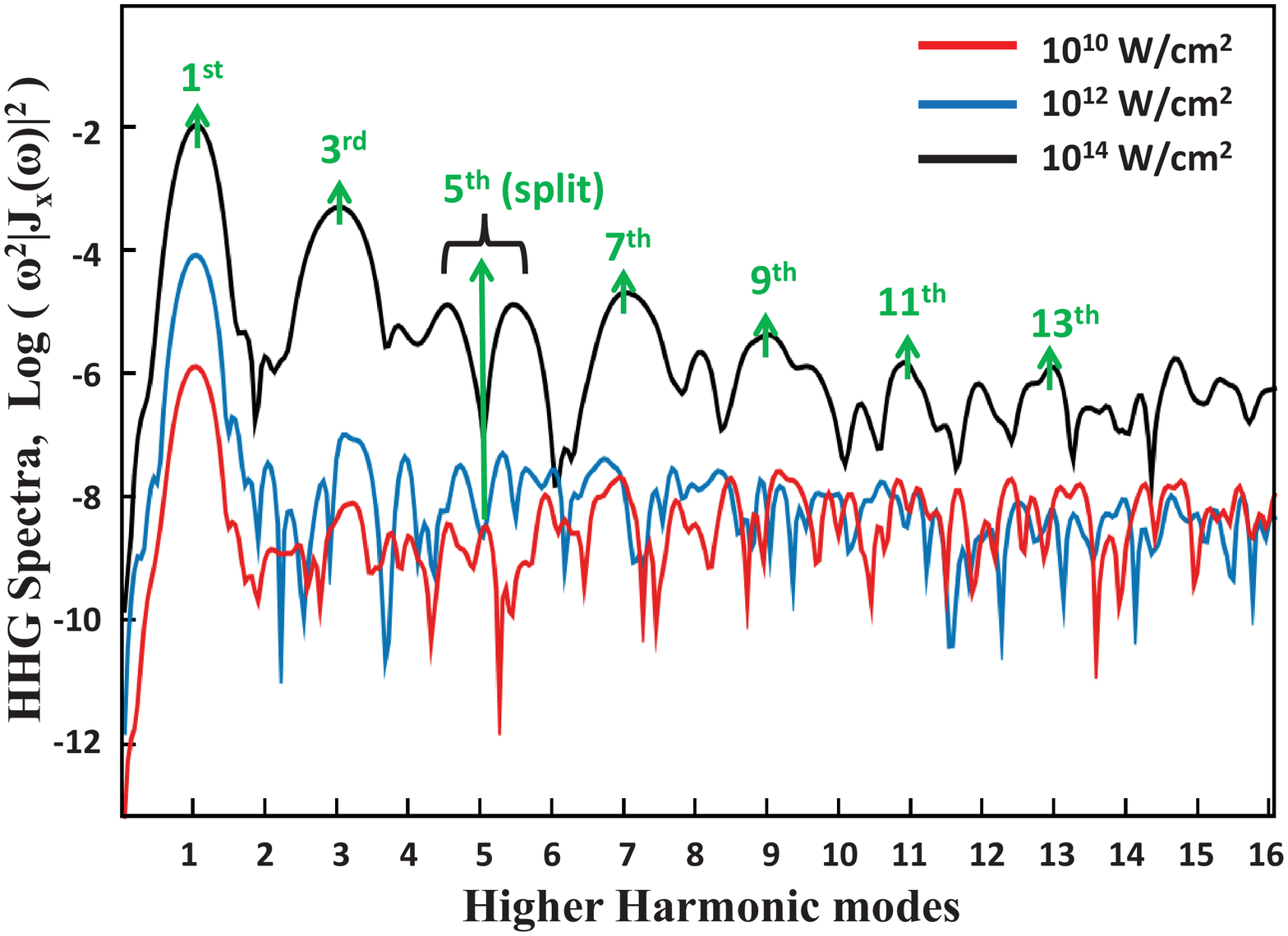}
  \caption
   {(Color online) HHG spectra calculated for the non-CDW phase of monolayer NbSe$_2$ for three different pulse intensities.
}\label{fig2}
\end{figure}

\begin{figure}
 \includegraphics[scale=0.34,angle=0]{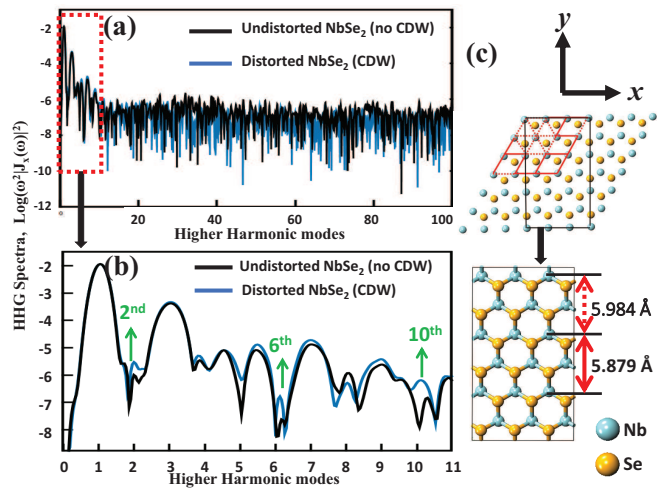}
  \caption
   {(Color online) (a) Comparing HHG spectra in the CDW distorted phase (blue) vs the non-CDW phase (black) of monolayer NbSe$_2$. The applied external field is along the (100)-direction. (b) The first 10 modes of the HHG spectra, as shown by the red dashed box in (a), are zoomed in here. Difference between the CDW (blue) and non-CDW (black) curves are shown for the first 10 modes of the HHG spectra. 
   (c) 2D NbSe$_2$ in the x-y plane. The top panel shows a 3x3 unit cell with red lines, 
   which is prepared according to the CDW ordering observed in earlier experimental work.\cite{NbSe2_1,NbSe2_3} The bottom panel shows the conventional unit cell taken for this work. 
}\label{fig3}
\end{figure}

\subsection{\label{sec:level2b}Computational Approach}
The above formalism for RT-TDDFT has been implemented in a handful of codes such as TDAP~\cite{tdap}, RT-SIESTA~\cite{velocity_gauge}, OCTOPUS~\cite{etrs}, and SALMON~\cite{salmon_1}, among a few others, with some variations to adapt to their original DFT methods. We used SALMON for the results presented here, which showed efficient scalable behavior in a modern HPC platform, particularly suitable for our systems of interest. Since SALMON can only treat rectangular unit cells, we considered a rectangular cut (Fig.~3(c)) with a 3$\times$6 supercell of 2H-NbSe$_2$, which is the minimal size for capturing the experimentally established~\cite{NbSe2_1,NbSe2_3} CDW distorted periodic lattice. We used 2$\times$2$\times$2 $k$-point sampling in the 
Brillouin zone (BZ) of the supercell. Higher accuracy was achieved using a 28$\times$32$\times$30 real-space grid. We have used  Trouiller-Martin type LDA-FHI pseudo-potentials for Nb and Se atoms~\cite{PP_1,PP_2}, where 5 ($4d^45s^1$) and 6 ($4s^24p^4$) valence electrons were considered, respectively. 

For easier comparison with experiments, we have used a femtosecond laser pulse polarized along the [100] crystal axis ($x$-direction in xy plane) with a pulse shape whose x-component is defined as: 

\begin{eqnarray}
\label{eqn:eqn_09}
A_x(t)=-c\frac{E_o}{\omega_o}\cos(\omega_o t)\sin^2(\frac{\pi t}{\tau_L})\Theta(\tau - t),
\end{eqnarray} 
where the electric field can alternatively be defined as: 
\begin{eqnarray}
\label{eqn:eqn_10}
E_x(t) &=& -\frac{1}{c}\frac{\partial A_x(t)}{\partial t} \nonumber \\
&=& \frac{E_o}{\omega_o}\frac{\partial}{\partial t}\left[\cos(\omega_o t)\sin^2(\frac{\pi t}{\tau_L}) \Theta(\tau - t)\right ].
\end{eqnarray} 

In this calculation, we have used a pulse-width of $\tau$=30 fs and photon energy $\omega$ = 0.6 eV. The intensity of the pulse is related to the amplitude by $I$ $\approx$ c$E^2_o$/8$\pi$. This equality only holds for infinite unmodulated wave trains (not pulses), but we ignored these subtle differences and considered the above approximate relation, following ref:~\onlinecite{Yabana_2}.
The external field ($E_x(t)$) pulse shapes for three different intensities ($10^{10}$, $10^{12}$, and $10^{14}$ W/cm$^2$) are presented in Fig.~1(a).  The Fourier transformed components (real and imaginary) of the most intense pulse ($10^{14}$ W/cm$^2$) are shown in Fig.~1(b). In this Fourier transformed plot, the dominant peak is observed around $\omega$ = 0.6 eV. 
The electronic response is also expected to be strongest at this frequency, a.k.a. the fundamental frequency, as discussed below in the `results' section. 

Calculations for both CDW and non-CDW structures were performed for a 55 fs time-propagation using $\Delta t$ = 0.0019 fs as the time step. Such small time-steps are required for the computational stability of our real-time propagation algorithm. Both these values were tested to be sufficient for obtaining a converged HHG spectra. For time-propagation, we used an `enforced time-reversal symmetry' propagator~\cite{etrs}. 

\section{\label{sec:level3}Results and Discussion}
\begin{figure}
 \includegraphics[scale=0.34,angle=0]{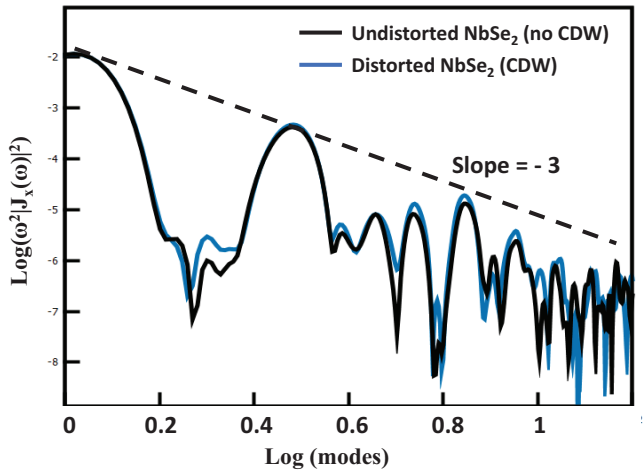}
  \caption
   {(Color online) Log-log plot for HHG signal vs mode number. The linear fit shows the power law decay of the peak strengths of the HHG spectra for both CDW (blue) and non-CDW (black) structural phases of NbSe$_2$. 
}\label{fig4}
\end{figure}
One of our primary motivations is to identify distinct features in the HHG spectra when comparing CDW and non-CDW ordered phases of monolayer 2H-NbSe$_2$ under strong optical excitation. This helps establish a correlation between the external intensity and the nonlinear electronic response. Starting with the non-CDW or undistorted structural phase and using the RT-TDDFT approach, we calculated time-dependent current densities at three different intensities: 10$^{10}$, 10$^{12}$, and 10$^{14}$ W/cm$^2$. The dominant peak features in the $J(t)$ spectra are found up to 30 fs, which is also the width of the laser pulse. Beyond 30 fs, the tail of the spectra shows a complex modulated pattern with finer spectral features, as shown in the inset of Fig.~1(c), for an intensity of 10$^{14}$ W/cm$^2$. This is a robust signature of the nonlinear response of the system. 

Under intense optical excitation, the conduction and/or valence electrons undergo both interband and intraband transitions. The spectral features of the real-time current density take the shape of the incident laser pulse, which is an indicator of conduction electron response in metallic monolayer NbSe$_2$. At high enough intensities, current density spectra (inset of Fig.1(c)) capture the nonlinear response generated due to both inter- and intraband transitions. Such features are best represented in the Fourier transformed frequency space, where various higher order modes in the current density spectra are readily identified. Since our pulse frequency was set to 0.6 eV, we represented the odd and even modes as the odd and even integer multiples of this fundamental frequency, respectively, for the HHG spectra.  

\begin{figure}
 \includegraphics[scale=0.34,angle=0]{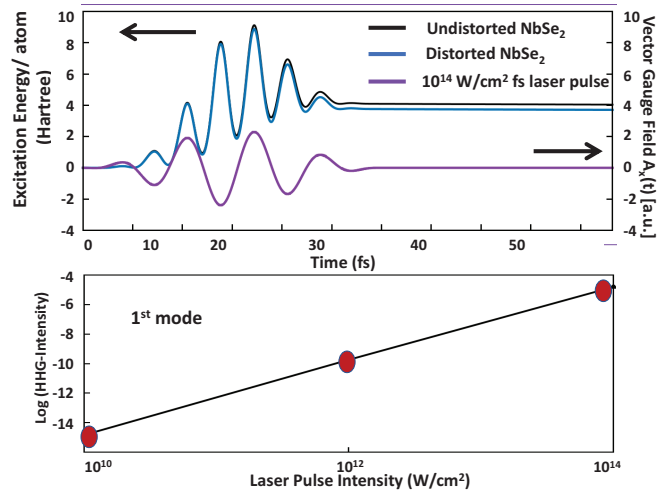}
  \caption
   {(Color online) Top panel: Excitation energy as a function of time for CDW (blue) and non-CDW (black) ordered 2D NbSe$_2$. The laser field with strongest intensity of $10^{14}$ W/cm$^2$ is the purple curve. Bottom panel: Log plot of HHG intensity against laser field intensity. The slope of the linear fit represents the power of the scaling behavior (see text).
}\label{fig5}
\end{figure}
The nonlinear high harmonic response of materials is symmetry dependent, and the corresponding optical selection rules~\cite{boyd} prohibit any even modes from appearing when the system is purely centrosymmetric with full inversion symmetry. However, the monolayer 2H-NbSe$_2$ crystal structure is noncentrosymmetric (i.e., inversion symmetry is broken). Consequently, the HHG spectra show even modes besides the stronger odd mode peaks (Fig.~2). Similar even modes were also observed in the HHG spectrum of monolayer MoS$_2$\cite{even-mode} which were argued to be due to intra-band currents\cite{Ghimire2019} originating from valley-contrasting Berry curvature in the absence of inversion symmetry.\cite{berry1,berry2} For all the three pulse intensities considered in this work, the first or fundamental peak is strongest and increases with laser intensity. However, the rest of the higher harmonic mode peaks become prominent only for the highest laser intensity $10^{14}$ W/cm$^2$ (solid black spectra in Fig.~2), where almost no enhancement was noticed for $10^{10}$ and $10^{12}$ W/cm$^2$ laser pulses. For these lower intensities, the HHG spectra reach a plateau right after the fundamental peak, with spectra that appear noisy for higher harmonics. On the other hand, several odd and even harmonic peaks become distinct for an intensity of $10^{14}$ W/cm$^2$. 

One naturally wonders if HHG spectra have the
ability to identify features related to various 
photoinduced structural phases of matter. For verification, we prepared a CDW ordered 3$\times$6 lattice structure following earlier experimental work\cite{NbSe2_1,NbSe2_3} (Fig.~3c), and computed the HHG spectra for an intensity of $10^{14}$ W/cm$^2$. Interestingly, enhancements of the 2nd, 6th and 10th harmonic modes were observed (Fig.~3b) for the distorted CDW phase (solid blue) in comparison to the non-CDW phase (solid black) of 2D NbSe$_2$. This suggests that HHG spectra can provide a useful probe for structural phases, such as CDW ordering in low dimensional systems. 

The electronic response of materials under strong laser fields manifests itself through different power law scaling behavior of the HHG spectra. 
The peak heights of the HHG spectrum for the highest intensity (solid black curve in Fig.~2) clearly show a power law decay with increasing mode number, which is another signature of the electronic response of the system\cite{power-law1} under strong driving field. This power law decay of the peak intensity is $\approx$ -3, as shown in Fig.~4 with a log-log plot for both the CDW (solid blue) and non-CDW NbSe$_2$ (solid black).

Yet another power law behavior of HHG spectra is observed as a function of the intensity of the incident laser pulse. Both the frequency (mode) and intensity dependent behaviors are caused by various linear and nonlinear quantum processes, including tunneling and multi-photon transitions in a strong laser-field, as explained by the Keldysh theory of ionization.\cite{keldysh,power-law1}  We observed the intensity-dependence of the fundamental HHG peak as shown in Fig.~5 (lower panel). 

The intra- and inter-band transitions are the two key features of electronic response of materials. The excitation energy can provide some qualitative insights into such processes. In the top panel of Fig.~5, we have shown the time evolution of the excitation energy of both CDW (solid blue) and non-CDW (solid black) NbSe$_2$, plotted against the laser intensity of
$10^{14}$ W/cm$^2$. The oscillatory behavior of the excitation energy with time qualitatively shows intraband transitions of conduction electrons in metallic monolayer 2H-NbSe$_2$, which oscillate in their respective bands under the external laser field. On the other hand, the step-wise increment of the excitation energy (e.g., at 25 fs) qualitatively represents interband transitions (due to total energy increase). Here the energy increases as the valence electrons tunnel into the low lying unoccupied states under the external laser field. 

\section{Conclusion}

Our theoretical study of high harmonic generation in monolayer NbSe$_2$ under ultrashort pulsed optical excitation has demonstrated significant promise in identifying the nonlinear response and distinct HHG peak features between CDW and non-CDW structural phases. The characteristic features of the HHG spectral peaks show the underlying symmetry properties and how they may become impacted with increasing laser intensity. We have theoretically identified the enhancement of specific higher order even modes (e.g. the 2nd 6th, and 10th modes) in the HHG spectra of CDW ordered NbSe$_2$, in comparison to that of the non-CDW ordered phase. These predictions provide guidance for future experiments. Our study also confirms different power law scaling behaviors of HHG spectra, which are characteristic indicators of the nonlinear response of monolayer 2H-NbSe$_2$ system under a strong laser field. Finally, the intra- and inter-band transition processes are also revealed in our RT-TDDFT calculated excitation energy spectrum. The work presented here thus charts a general path forward for discovering new `tuning' and `control' principles for lower dimensional TMDC materials with competing quantum phases.  
\section{Acknowledgement}
We thank Benedikt Fauseweh, Prashant Padmanabhan, and Daniel Rehn for many valuable discussions.  
This   work   was   carried   out   under   the   auspices of  the  U.S.  Department  of  Energy  (DOE)  National Nuclear  Security  Administration  under  Contract  No. 89233218CNA000001 and was supported by the LANL LDRD Program under the Project No. 20190026DR.  We acknowledge the support by the  Institutional  Computing  Program at LANL and NERSC, via the Center for Integrated Nanotechnologies, a DOE BES user facility,  for computational resources.

\bibliography{references.bib}

\end{document}